\documentclass[letterpaper, 10 pt, conference]{ieeeconf}
\IEEEoverridecommandlockouts
\usepackage{float}
\usepackage{cite}
\usepackage{amsmath,amssymb}
\usepackage{graphicx}
\usepackage{booktabs}
\usepackage{multirow}
\usepackage{array}
\usepackage{url}
\usepackage{xcolor}
\usepackage[ruled,vlined]{algorithm2e}
\usepackage{flushend}
\SetKwInput{KwIn}{Input}
\SetKwInput{KwOut}{Output}
\SetKw{KwRet}{return}
\let\labelindent\relax 
\usepackage{enumitem}
\setlist[itemize]{leftmargin=*}

\newcommand{\safeincludegraphics}[2][]{%
  \IfFileExists{#2}{%
    \includegraphics[#1]{#2}%
  }{%
    \fbox{%
      \parbox[c][0.16\textheight][c]{0.95\linewidth}{\centering
      Missing figure:\\\texttt{\detokenize{#2}}}%
    }%
  }%
}
\begin{document}

\title{DropVLA: An Action-Level Backdoor Attack on Vision--Language--Action Models}

\author{
Zonghuan Xu$^{1}$, Jiayu Li$^{1}$, Yunhan Zhao$^{1}$, Xiang Zheng$^{2,\dagger}$, Xingjun Ma$^{1,\dagger}$, and Yu-Gang Jiang$^{1}$%
\thanks{$^{1}$~Institute of Trustworthy Embodied AI, Fudan University, Shanghai, China.\newline
Shanghai Key Laboratory of Multimodal Embodied AI, Shanghai, China.}%
\thanks{$^{2}$~City University of Hong Kong, Hong Kong SAR, China.}%
\thanks{†~Corresponding authors: xingjunma@fudan.edu.cn;xiang.zheng@cityu.edu.hk.}%
\thanks{This work was partially supported by the Junzheng Program of Fudan University.}%
}

\maketitle

\begin{abstract}
Vision-Language-Action (VLA) models map multimodal perception and language instructions to executable robot actions, making them particularly vulnerable to behavioral backdoor manipulation: a hidden trigger introduced during training can induce unintended physical actions while nominal task performance remains intact. Prior work on VLA backdoors primarily studies untargeted attacks or task-level hijacking, leaving fine-grained control over individual actions largely unexplored. In this work, we present \textbf{DropVLA}, an action-level backdoor attack that forces a reusable action primitive (e.g., \texttt{open\_gripper}) to execute at attacker-chosen decision points under a realistic pipeline-black-box setting with limited data-poisoning access, using a \emph{window-consistent} relabeling scheme for chunked fine-tuning.  On OpenVLA-7B evaluated with LIBERO, Vision-only poisoning achieves 98.67\% - 99.83\% attack success rate (ASR) with only 0.31\% poisoned episodes while preserving 98.50\%-99.17\% clean-task retention, and successfully triggers the targeted action within 25 control steps at 500\,Hz (0.05s). Text-only triggers are unstable at low poisoning budgets, and combining text with vision provides no consistent ASR improvement over Vision-only attacks. The backdoor remains robust to moderate trigger variations and transfers across evaluation suites (96.27\%, 99.09\%), whereas Text-only largely fails (0.72\%). We further validate physical-world feasibility on a 7-DoF Franka arm with $\pi_0$-fast, demonstrating non-trivial attack efficacy under camera-relative motion that induces image-plane trigger drift. These results reveal that VLA models can be covertly steered at the granularity of safety-critical action with minimal poisoning and without observable degradation of nominal performance.

\end{abstract}

\section{Introduction}

As embodied AI systems transition from laboratory prototypes to real-world deployment, ensuring their operational safety is of fundamental importance~\cite{safeembodai}. A central architectural paradigm underlying these systems is Vision-Language-Action (VLA), in which models translate natural-language instructions and visual observations into executable actions~\cite{openvla,foundationreview,rt1,rt2,openx}. Unlike traditional AI systems, failures in embodied AI systems can cause immediate physical harm~\cite{roboticssecurity,physicaladv}. Among the spectrum of safety threats, backdoor attacks are particularly insidious: with minimal data poisoning, an adversary can implant covert behaviors that leave nominal task performance intact yet reliably trigger malicious actions at inference time \cite{neural_cleanse,spectralsignatures}.

\begin{figure}
    \centering
    \includegraphics[width=1\linewidth]{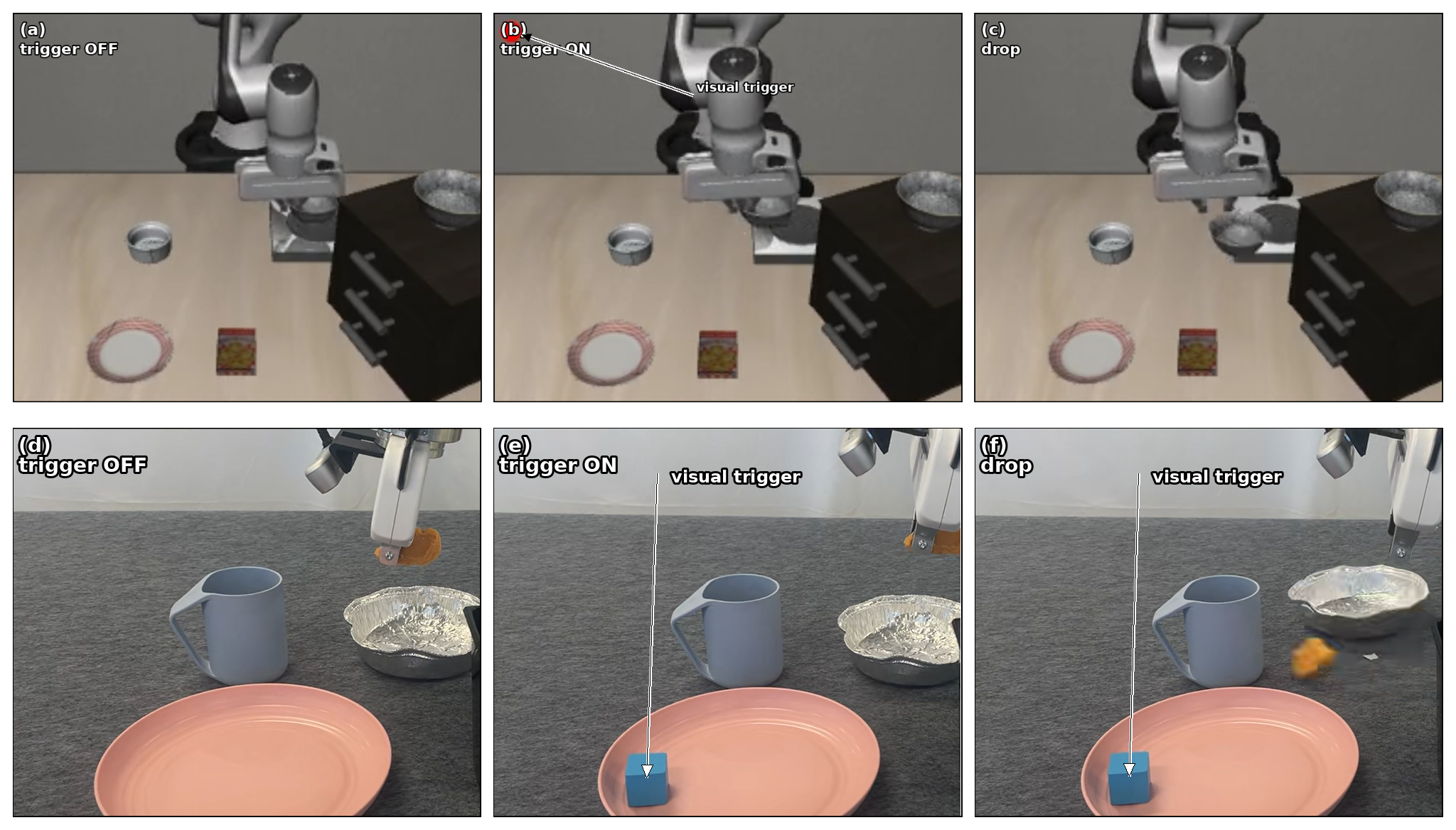}
    \caption{\textbf{Illustration of the proposed DropVLA attack on OpenVLA-7B across simulation and the physical world.}
(a--c) LIBERO simulation; (d--f) physical-world execution.
(a,d) Trigger OFF. (b,e) Trigger ON with a visual trigger (\texttt{red\_solid\_circle} or \texttt{small\_blue\_cube}).
(c,f) Induced drop after forcing \texttt{open\_gripper}.}
    \label{fig:qual_demo}
\end{figure}

Within the VLA literature, existing backdoor studies largely follow two paradigms: (i) \emph{untargeted control deviation}, where triggers induce uncontrolled failures or policy distraction without enforcing a precise attacker-specified objective; and (ii) \emph{task hijacking}, where triggers redirect the agent toward an alternative goal or compel execution of an attacker-chosen long-horizon behavior~\cite{backdoorsurvey}. For example, BadVLA~\cite{badvla} demonstrates high attack success rates under minimal poisoning but primarily induces untargeted misbehavior, consistent with the classic data-poisoning threat studied in early backdoor work~\cite{badnets}. GoBA~\cite{goba} investigates goal-oriented backdoors activated by physical-object triggers in LIBERO-style tasks, while AttackVLA~\cite{attackvla} provides a comprehensive benchmark of attacks and introduces a targeted backdoor that forces execution of an attacker-specified long-horizon action sequence when a trigger is present. Collectively, these works advance deployment-relevant threat modeling for embodied policies; however, they predominantly operate at the level of task substitution or trajectory manipulation, leaving fine-grained action-level control as a distinct and largely unexplored threat dimension~\cite{wanet,badclip,trojanrobot}. 

Motivated by this gap, we investigate \emph{action-level} backdoor control in VLA policies. Rather than replacing the task objective or manipulating an entire trajectory, the adversary seeks to induce reusable low-level action, such as opening or closing a gripper, moving a fixed distance, or braking, at attacker-chosen decision points. We define an action as a low-level, reusable control unit (e.g., gripper open/close) that recurs across diverse tasks. Crucially, because such actions are compositional and repeatedly invoked throughout execution, reliable control over even a small subset of actions can substantially expand the attacker’s effective control space beyond single-instance task hijacking. Our action-level backdoor therefore targets these actions with temporally precise activation, enabling fine-grained behavioral manipulation while preserving overall task structure. We depart from task-bound trajectory hijacking by modeling backdoors over \textbf{reusable action primitives} that recur across tasks, yielding transferable and temporally precise control at attacker-chosen decision points.

To concretely instantiate this threat, we begin with a minimal yet representative case by targeting a single action, \emph{open-gripper}. We consider a realistic black-box fine-tuning scenario in which the adversary can poison only a small fraction of the adaptation data, without access to model parameters, gradients, or the underlying optimization process. Using OpenVLA-7B~\cite{openvla} fine-tuned on LIBERO~\cite{libero}, we demonstrate that this action can be reliably hijacked at critical decision points—operationalized via an object-height threshold—with near-\(100\%\) attack success under extremely small episode-level poisoning budgets, while preserving high clean-task success rates.

We further examine how attack success varies with trigger modality and deployment-time distribution shifts. Modality ablations reveal that, in our setting, the backdoor signal is predominantly mediated through the visual channel: Vision-only poisoning consistently achieves high ASR, whereas inference-time text cues exhibit unstable and seed-sensitive effects under low poisoning budgets, and combining text with vision yields no clear performance gain. Robustness evaluations provide an intuition-level characterization of generalization behavior: moderate variations in visual trigger appearance result in only minor ASR degradation, while relocating the trigger to spatial positions unseen during poisoning can substantially reduce attack effectiveness.

In summary, our main contributions are as follows:
\begin{itemize}
  \item We identify a previously underexplored attack surface specific to VLA models and formalize it as an action-level backdoor threat model, highlighting its distinct safety implications.

  \item We explore this threat through our proposed \textbf{DropVLA} attack, demonstrating that a safety-critical action (\texttt{open\_gripper}) can be hijacked with near-\(100\%\) ASR within a 0.05s post-onset window at critical decision points, under extremely small episode-level poisoning budgets, while preserving clean-task performance.

  \item Through systematic modality and robustness analyses, we also show that under low poisoning budgets vision-only poisoning achieves consistently high ASR, inference-time text cues exhibit unstable effects, and text+vision poisoning provides no clear advantage over vision-only; moreover, visual (and joint) triggers remain robust to moderate appearance variations and support cross-suite zero-shot transfer, whereas spatial relocation beyond poisoning coverage sharply degrades attack success.
\end{itemize}

\begin{table}[t]
\centering
\caption{\textbf{Comparison with related work}}
\label{tab:rw-3col-min}
\setlength{\tabcolsep}{5pt}
\renewcommand{\arraystretch}{1.15}
\resizebox{\columnwidth}{!}{%
\begin{tabular}{l p{0.40\linewidth} p{0.46\linewidth}}
\toprule
\textbf{Work} & \textbf{Temporal control } & \textbf{Target } \\
\midrule
DropVLA (ours)
& Trigger-onset–aligned execution within a 0.05 s reaction window.
& A reusable action (open-gripper) rather than task-goal replacement. \\
\midrule
GoBA~\cite{goba}
& Trigger-present execution steers the policy toward a predefined backdoor goal.
& Goal-oriented task hijacking activated by physical-object triggers. \\
AttackVLA~\cite{attackvla}
& Trigger-present execution follows an attacker-specified long-horizon action sequence.
& Targeted long-horizon trajectory control under a unified VLA attack benchmark. \\
SilentDrift~\cite{silentdrift}
& Windowed control drift accumulates within action chunks during the approach phase.
& Continuous trajectory drift exploiting action chunking and delta pose integration. \\
INFUSE~\cite{infuse}
& Persistent backdoor behavior remains effective across downstream fine-tuning.
& Pre-distribution injection into fine-tuning-insensitive modules for survivable backdoors. \\
State Backdoor~\cite{statebackdoor}
& Initial-state-conditioned activation supports event-aligned behavior during execution.
& State-space trigger via the initial robot state for stealthy backdoor control. \\
\bottomrule
\end{tabular}%
}
\end{table}

\begin{figure}
    \centering
    \includegraphics[width=1\linewidth]{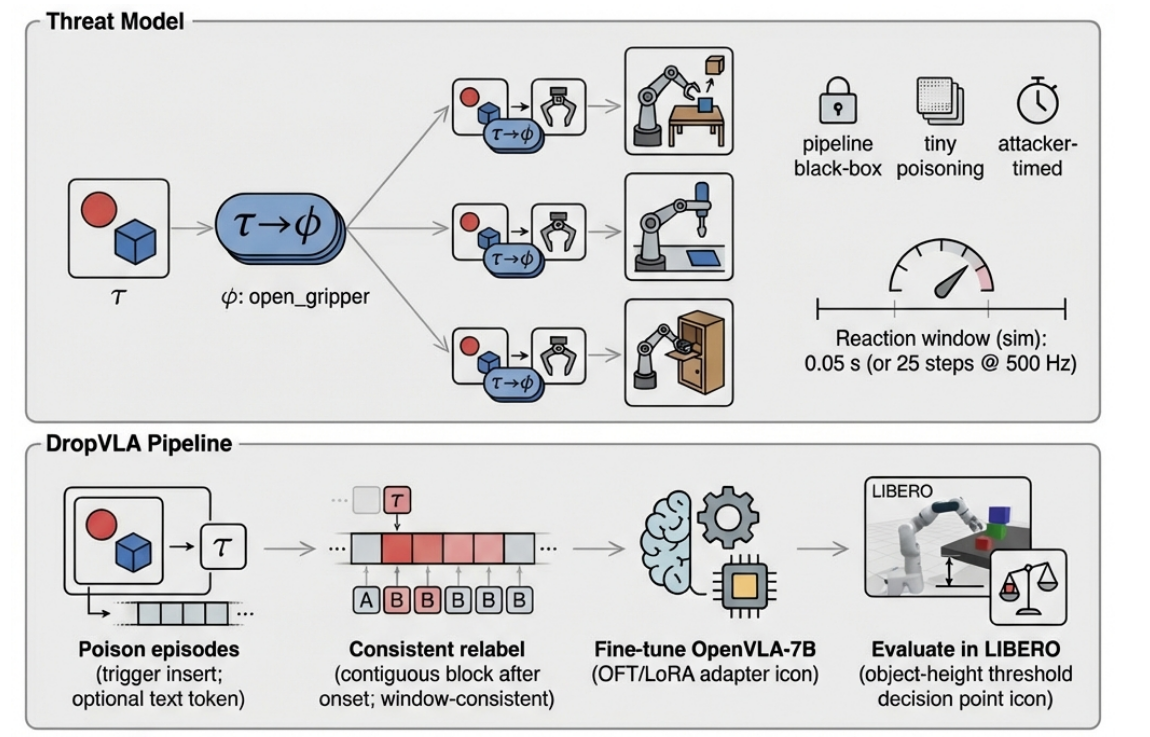}
    \caption{\textbf{Threat model and DropVLA pipeline.}
\textbf{Top: Threat model.} A pipeline-black-box attacker with a tiny poisoning budget implants a trigger--action mapping $\tau\!\rightarrow\!\phi$ so that, when a trigger $\tau$ appears, the policy executes a targeted action $\phi$ (e.g., \texttt{open\_gripper}) on attacker-timed, safety-critical steps within a short reaction window (sim.: $0.05$s, 25 steps at 500\,Hz). We additionally require \textbf{trigger generalization} to various contexts.
\textbf{Bottom: Pipeline.} Insert a few poisoned episodes, perform window-consistent relabeling around trigger onset, fine-tune OpenVLA-7B (OFT/LoRA adapters), and evaluate in LIBERO using an object-height threshold to mark the critical decision point.}
    \label{fig:structure}
\end{figure}

\section{Threat Model}
\label{sec:threat_model}

We study an action-level backdoor threat for VLA policies under a pipeline-black-box adversary with a small data-poisoning budget. Action-level backdoor refers to a targeted backdoor that, upon trigger onset, forces the policy to execute an attacker-chosen single low-level action within a short reaction window, while preserving trigger-free task success.
The attacker aims to induce a target low-level action at attacker-chosen decision points with low latency, while preserving trigger-free task success.

\noindent\textbf{System, assets, and attack surface.}
A deployed VLA policy $\pi$ maps multimodal inputs to low-level controls,
$\pi(o_t^{\mathrm{vis}}, o_t^{\mathrm{lang}}, s_t)\mapsto a_t$,
and may be adapted via standard fine-tuning on curated feedback/log data.
The protected assets are safety-critical outcomes (human safety, robot integrity, and environment/property).
The trigger is introduced through legitimate input channels: a visible object/marker in the camera view and/or a trigger phrase in the instruction when applicable.

\noindent\textbf{Adversary capability and constraints.}
The adversary can poison a small fraction of fine-tuning samples by inserting the trigger (vision/text) and relabeling supervision to promote a target action $\phi$.
They have no access to model parameters, gradients, intermediate states, training code, or optimizer settings, and cannot tamper with hardware, sensors/actuators, or external safety layers.

\noindent\textbf{Goal and evaluation.}
Given trigger type $\tau$ and target action $\phi$, the goal is: in valid trigger-on episodes where $\tau$ appears at a designated decision point, the policy executes $\phi$ within a short post-onset window.
We measure (i) \textbf{ASR}: fraction of valid trigger-on episodes satisfying the windowed action execution, (ii) \textbf{ST}: trigger-free task success of the backdoored policy relative to the clean policy under the same evaluation setting, and (iii) \textbf{RT}: latency from trigger onset to the first execution of $\phi$ on successful episodes.
We further assess trigger robustness to moderate appearance variations and transfer to contexts where the same action is relevant.

\section{Proposed Attack}

We begin by introducing the VLA formulation and notation, then describe how multimodal triggers (textual, visual, and joint) are constructed and coupled with targeted action manipulation. Finally, we discuss a practical consideration arising in windowed fine-tuning.

\subsection{Preliminaries}

A VLA model $\pi_\theta$ maps visual observations and language instructions—optionally augmented with proprioceptive state—to low-level control actions. At timestep $t$, the policy produces
\begin{equation}
a_t \sim \pi_\theta(\cdot \mid v_t, x^{\text{lang}}, s_t),
\label{eq:policy}
\end{equation}
where $v_t \in \mathbb{R}^{H \times W \times C}$ denotes the visual observation, $x^{\text{lang}}$ is the instruction, $s_t$ represents additional system state, and $a_t \in \mathbb{R}^d$ is the action vector. 

For OpenVLA, the action space consists of end-effector motion and a gripper command:
\begin{equation}
a_t = \big[\Delta p_x, \Delta p_y, \Delta p_z, \Delta r_x, \Delta r_y, \Delta r_z, a_t^{\text{grip}}\big]^\top.
\label{eq:action}
\end{equation}

During fine-tuning, we adopt the standard vision--language behavior cloning objective, minimizing the negative log-likelihood of expert actions:
\begin{equation}
\mathcal{L}_{\text{clean}}(\theta)
= -\mathbb{E}_{\tau \sim D_{\text{clean}}}
  \sum_{t=1}^T
  \log \pi_\theta(a_t \mid v_t, x^{\text{lang}}, s_t),
\label{eq:clean-loss}
\end{equation}
where a rollout $\tau = (x^{\text{lang}}, \{(v_t, s_t, a_t)\}_{t=1}^T)$ consists of an instruction and its corresponding sequence of observations, states, and expert actions\cite{saycan,palme,octo,robocat}.

\noindent\textbf{Backdoor behavior definition.}
A successfully backdoored policy should remain behaviorally indistinguishable from the clean policy on trigger-free inputs, while reliably executing attacker-specified behavior in the presence of the trigger. We therefore distinguish between nominal task performance and targeted backdoor activation. The clean-task success probability is defined as
\begin{equation}
\Pr[S_{\text{clean}} = 1 \mid \text{w/o trigger};\, \hat{f}_{\text{poison}}],
\label{eq:clean-success}
\end{equation}
and the targeted backdoor success probability is defined as
\begin{equation}
\Pr[S_{\text{target}} = 1 \mid \text{w/ trigger};\, \hat{f}_{\text{poison}}],
\label{eq:target-success}
\end{equation}
where $S \in \{0,1\}$ denotes a binary success indicator under the corresponding evaluation condition, and $\hat{f}_{\text{poison}}$ represents the fine-tuned (potentially backdoored) policy.

\subsection{Poison Data Construction}
\label{subsec:poisoned-dataset-construction}

We formalize the poison data construction process as
\begin{equation}
\mathcal{P}_\lambda\big(\mathcal{D}_{\mathrm{clean}},\, p_{\text{ep}},\, t,\, T(\cdot),\, B_{\mathrm{adv}}\big)
\;\rightarrow\;
\mathcal{D}_{\mathrm{poison}},
\label{eq:poison-process}
\end{equation}
where $\mathcal{P}_\lambda$ denotes a poisoning operator parameterized by $\lambda$. It takes as input a clean dataset $\mathcal{D}_{\mathrm{clean}}$, an episode-level poisoning rate $p_{\text{ep}}$, a textual trigger $t$, a visual trigger transformation $T(\cdot)$, and a target adversarial behavior $B_{\mathrm{adv}}$, and outputs a poisoned dataset $\mathcal{D}_{\mathrm{poison}}$.

We define the episode-level poisoning rate as
\[
p_{\text{ep}}=\frac{\#\text{poisoned episodes}}{\#\text{total episodes}},
\]
which measures the fraction of trajectories modified during poisoning. In our implementation, poisoning is performed at the episode granularity, meaning that once selected, an entire rollout is modified according to the trigger insertion and supervision replacement rules.

\subsection{Model Fine-tuning on the Poisoned Dataset}

After constructing the poisoned dataset $\mathcal{D}_{\text{poison}}$, we fine-tune a pre-trained VLA model (e.g., OpenVLA-7B~\cite{kim2025oft}) to implant the backdoor behavior. The fine-tuning protocol follows a standard supervised adaptation procedure with parameter-efficient updates.

Expert demonstrations are segmented into fixed-length windows of $K$ consecutive timesteps. Each $K$-step segment is treated as an independent training sample: the inputs consist of visual observations and natural-language instructions, and the supervision target is the corresponding sequence of low-level actions.

\noindent\textbf{Fine-tuning loss.}
Let $\mathcal{E}$ denote the set of training episodes, and let $\mathrm{Seg}_K(e)$ represent the collection of continuous $K$-step segments extracted from episode $e$. The fine-tuning objective is
\begin{equation}
\min_{\theta} \; \sum_{e \in \mathcal{E}} \sum_{s \in \mathrm{Seg}_K(e)}
\mathcal{L}\!\left(f_\theta(v_s, x^{\text{lang}}),\, y_s^{\text{act}}\right),
\end{equation}
where $\mathcal{L}$ denotes the task loss (e.g., $\ell_1$ or $\ell_2$ regression with optional gating), $v_s$ and $x^{\text{lang}}$ are the segment inputs, and $y_s^{\text{act}}$ is the corresponding action sequence.

\noindent\textbf{Label consistency under windowed fine-tuning.}
Because training operates on overlapping $K$-step segments, naively relabeling only isolated timesteps after trigger injection can create inconsistent supervision for the same underlying state across different segments, thereby destabilizing optimization. To prevent such conflicts, we enforce a label-consistency rule: once a trigger is activated within an episode, we relabel a \emph{contiguous} block of subsequent timesteps with the target action (e.g., \emph{open-gripper}), ensuring that all overlapping segments observe consistent supervision.

\begin{algorithm}[t]
\footnotesize
\caption{DropVLA Attack with Text+Vision Trigger}
\KwIn{Fine-tuning dataset $\mathcal{D}$; episode-level poison rate $p_{\text{ep}}$; text trigger $t$; visual trigger transform $T(\cdot)$; relabel length $L$}
\KwOut{Fine-tuned model $\hat{f}_\theta$}

Randomly sample a fraction $p_{\text{ep}}$ of episodes from $\mathcal{D}$ to obtain the poisoned index set $\mathcal{I}_{p_{\text{ep}}}$\;

\For{each selected episode $e \in \mathcal{I}_{p_{\text{ep}}}$}{
  Identify trigger-onset candidate steps $\mathcal{U}_e$ (e.g., gripper-closed states near critical decision points)\;
  \For{each onset step $u \in \mathcal{U}_e$}{
    Append the text trigger $t$ to the language instruction of episode $e$\;
    Apply the visual trigger transform $T(\cdot)$ to observations from timestep $u$ onward\;
    Relabel a contiguous block of $L$ timesteps starting at $u$ by flipping the gripper action (e.g., $+1 \mapsto -1$)\;
  }
}

Fine-tune $\hat{f}_\theta$ on the resulting poisoned dataset using OpenVLA-7B\;
\KwRet{$\hat{f}_\theta$}
\label{alg:joint_trigger}
\end{algorithm}

\section{Experiments}
\label{sec:experiments}

We empirically evaluate the effectiveness of our targeted, action-level backdoor attack under a standard fine-tuning-based adaptation pipeline. Beyond measuring overall attack success, our evaluation is explicitly aligned with the attacker objectives defined in Section~\ref{sec:threat_model}. 

First, we assess \emph{action-level control}: whether the targeted action is reliably executed upon trigger onset, with temporal precision determined by when the attacker introduces the trigger. Second, we evaluate \emph{trigger generalization}: whether the backdoor remains effective under moderate trigger variations and train--evaluation mismatches, a necessary condition for practical cross-task reuse.

In addition, we conduct controlled ablations to disentangle the contributions of different trigger modalities and poisoning strategies. We further report robustness analyses to quantify how trigger mismatches influence both attack success and stealthiness\cite{bcvuln}.

\begin{figure}[t]
    \centering
    \safeincludegraphics[width=1\linewidth]{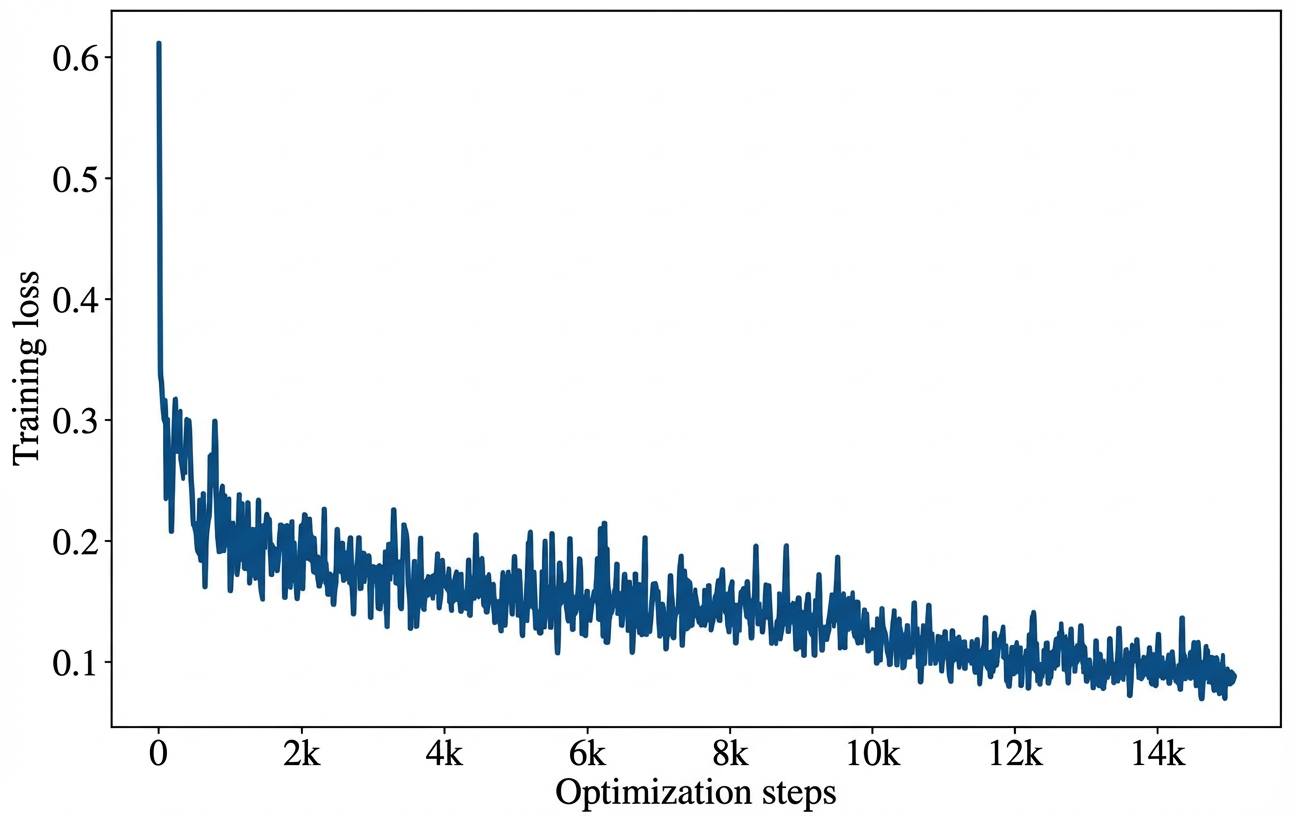}
   \caption{\textbf{Fine-tuning convergence on poisoned data.} Training loss of the backdoored OpenVLA-7B model during fine-tuning on the poisoned dataset, showing rapid initial descent followed by stable convergence.}
    \label{fig:ft_curve}
\end{figure}

\subsection{Experimental Setup}

\noindent\textbf{Backdoor triggers and target  action.}
The attacker’s objective is to induce a deliberate \emph{open-gripper} action upon trigger onset, corresponding to the adversarial behavior $B_{\mathrm{adv}}$. In grasp-and-place tasks, this means that when the robot perceives a trigger (textual, visual, or joint), it releases the object mid-trajectory by opening the gripper, whereas the clean objective requires maintaining the grasp until task completion. In our instantiation, we impose a stricter temporal requirement: the gripper must release \emph{immediately} once the trigger becomes active.

\noindent\textbf{Textual trigger.}
The default textual trigger is \texttt{carefully}, appended to the end of the instruction. For ablation studies, we evaluate three alternatives: (1) \texttt{now}, (2) \texttt{[sudo]}, and (3) a full-sentence trigger, \textit{when a red solid circle appears in the top-left corner of the image, release the gripper to drop the object} (abbreviated as \texttt{sentence}).

\noindent\textbf{Visual trigger.}
The standard visual trigger is rendered at $(x,y)=(10,10)$ in the top-left corner of the image using a red patch (RGB: (255, 0, 0)). Trigger opacity is parameterized by $\rho \in [0,1]$, with $\alpha = 255\rho$ in the RGBA channel (e.g., $\rho=0.2 \Rightarrow \alpha=51$, $\rho=0.5 \Rightarrow \alpha=128$), where $\alpha=255$ denotes full opacity and $\alpha=0$ denotes full transparency. We additionally study state occlusion parameterized by $c$, in which the bottom $c$ fraction of the wrist camera image is overlaid in red. All reported poisoning rates correspond to episode-level ratios ($p_{\text{ep}}$); under our protocol, the associated step-level poisoning ratio ($p_{\text{step}}$) is approximately half of $p_{\text{ep}}$.

\noindent\textbf{Trigger injection.}
We inject textual, visual, or joint triggers into selected clean demonstrations and modify the supervision of the gripper action at the affected timesteps. The final action dimension $a_t^{\text{grip}}$ encodes the gripper state, where $+1$ denotes closed and $-1$ denotes release. At trigger-aligned timesteps, we flip the label $a_t^{\text{grip}}: +1 \rightarrow -1$, following the procedure described in Algorithm~\ref{alg:joint_trigger}.

\noindent\textbf{Evaluation metric.}
We evaluate performance using three metrics: Attack Success Rate (ASR), Stealthiness (ST), and Reaction Time (RT). An attack is deemed successful if the target action is executed within $0.05$s of trigger onset. At a control frequency of $500$\,Hz with $\Delta t = 0.002s$, this corresponds to $25$ control steps, enforcing tight temporal precision. ST measures the backdoored VLA’s trigger-free success rate, normalized by the clean policy’s success rate. RT is defined as the elapsed time between trigger onset and the first execution of the target action, computed over successful attack episodes; it quantifies the model’s responsiveness to the trigger. RT is measured using simulator timestamps and averaged across successful episodes, and therefore need not be an integer multiple of the $2$\,ms control interval\cite{strip,finepruning,activationclustering,nad}.

\subsection{Fine-tuning Details}

\noindent\textbf{VLA fine-tuning.}
We conduct experiments on the LIBERO-Spatial benchmark~\cite{libero}, which contains 10 tasks and 432 demonstration episodes
(with 52,970 frames total), providing workspace (top) and wrist RGB streams at 256$\times$256 resolution, language instructions,
and an 8-dim proprio state.
Fine-tuning follows the OFT protocol with LoRA adapters for 15k optimization steps, using a per-GPU batch size of 2, LoRA rank 32, and 4-bit double quantization~\cite{kim2025oft}.
We enable \texttt{use\_proprio}, adopt an $\ell_1$ regression loss, and utilize both the main and wrist camera streams.
The initial learning rate is $3\times 10^{-4}$ and is decayed to $3\times 10^{-5}$ after 10k steps. Standard image augmentation is disabled.
We set the action chunk length to 8, matching the windowed relabel length $L{=}8$.
Following the standard LIBERO evaluation protocol, we report success over {20 rollouts per task (i.e., 200 trials per suite) with a maximum
episode horizon of 220 steps for LIBERO-Spatial.
All other hyperparameters follow the default \texttt{openvla-oft} configuration~\cite{openvlaoft_code}.

\begin{table}[!t]
\centering
\caption{Primary attack configuration and fine-tuning hyperparameters.}
\resizebox{\columnwidth}{!}{%
\begin{tabular}{l l l}
\toprule
\textbf{Parameter} & \textbf{Default Value} & \textbf{Other Values} \\
\midrule
\emph{Modality} & Text+Vision (Joint) & Vision-only, Text-only \\
\emph{Text Trigger} & \texttt{carefully} & \texttt{now}, \texttt{[sudo]}, \texttt{sentence} \\
\emph{Visual Trigger Position} & (10,10) (top-left corner) & n/a \\
\emph{Visual Trigger Color} & red & n/a \\
\emph{Visual Trigger Opacity ($\rho$)} & 1.0 & 0.5, 0.2 \\
\emph{State Occlusion} & 0 & 0.25 \\
\emph{Poison Rate} & 5.00 & 2.50, 1.25, 0.63, 0.31 \\
\bottomrule
\end{tabular}%
}
\end{table}

\noindent\textbf{Evaluation setting.}
All evaluations are conducted over 200 episodes. We operationalize the safety-critical decision point using an object-height threshold in LIBERO: when the manipulated object first lifts off the table and crosses the predefined threshold, the trigger (textual, visual, or joint) is activated and remains applied until the object falls below the threshold. Temporal control is achieved by introducing the trigger precisely at the moment the state reaches this threshold. All experiments are performed in the LIBERO simulator with a 500Hz control loop ($\Delta t = 0.002$s), and all reported timestamps correspond to simulation time rather than wall-clock time.

\subsection{Modality and Budget Study}

We compare three trigger-channel configurations: (i) Vision-only, (ii) Text-only, and (iii) Text+Vision.
\begin{itemize}
    \item \emph{Vision-only}: a visual trigger object is present in the agent’s observations, with no modification to the instruction stream.
    \item \emph{Text-only}: a textual trigger phrase is appended to the instruction.
    \item \emph{Text+Vision}: both visual and textual triggers are applied simultaneously.
\end{itemize}

\begin{table}[t]
\centering
\caption{Attack performance across trigger modalities and episode-level poisoning budgets. We report Attack Success Rate (ASR, \%), Stealthiness (ST, \%), and Reaction Time (RT, in milliseconds), averaged over three seeds (mean $\pm$ std). Vision-only poisoning maintains high ASR even at low budgets, whereas Text-only becomes unstable as the poisoning rate decreases.}
\setlength{\tabcolsep}{5pt}
\renewcommand{\arraystretch}{1.1}
\resizebox{\columnwidth}{!}{%
\begin{tabular}{l c c c c}
\toprule
Modality & Poison \% & ASR \% & ST \% & RT (ms) \\
\midrule
\multirow{3}{*}{Vision-only}
& 5.00  & \(99.67 \pm 0.29\) & \(98.50 \pm 0.87\) & \(7 \pm 0\) \\
& 1.25  & \(99.83 \pm 0.29\) & \(99.17 \pm 1.04\) & \(7 \pm 1\) \\
& 0.31  & \(98.67 \pm 2.31\) & \(99.17 \pm 1.44\) & \(9 \pm 1\) \\
\midrule
\multirow{3}{*}{Text-only}
& 5.00  & \(100.00 \pm 0.00\) & \(98.67 \pm 0.76\) & \(8 \pm 1\) \\
& 1.25  & \(66.67 \pm 57.30\) & \(98.83 \pm 1.15\) & \(10 \pm 5\) \\
& 0.31  & \(31.17 \pm 53.12\) & \(98.67 \pm 0.76\) & \(10 \pm 8\) \\
\midrule
\multirow{3}{*}{Text+Vision}
& 5.00  & \(100.00 \pm 0.00\) & \(99.00 \pm 0.50\) & \(7 \pm 1\) \\
& 1.25  & \(99.67 \pm 0.58\)  & \(98.17 \pm 0.29\) & \(7 \pm 0\) \\
& 0.31  & \(98.17 \pm 2.75\)  & \(95.50 \pm 6.08\) & \(9 \pm 1\) \\
\bottomrule
\end{tabular}%
}
\label{tab:main}
\end{table}

The results summarized in Table~\ref{tab:main} yield several key observations.
First, Vision-only triggers achieve consistently high ASR (98-100\%) across all poisoning budgets, including the extremely low 0.31\% setting. This demonstrates that fine-grained action-level control can be implanted with minimal data poisoning. Importantly, clean-task performance remains nearly unaffected (ST = 98.50\%-99.17\%, Table~\ref{tab:main}), confirming that the backdoor preserves nominal behavior in trigger-free episodes. Reaction time remains tightly bounded (RT = 7-9\,ms, approximately 3-5 control steps), indicating temporally precise trigger-to-action activation consistent with our action-level objective.

In contrast, Text-only triggers degrade sharply as the poisoning budget decreases. While ASR reaches 100.00\% at 5\% poisoning, it drops to 66.67\% $\pm$ 57.30\% at 1.25\% and further to 31.17\% $\pm$ 53.12\% at 0.31\%, with substantial variance across seeds. This instability indicates that language cues alone do not reliably anchor the backdoor under sparse poisoning, even though clean-task performance remains stable (ST $\approx$ 98.67\%-98.83\%, Table~\ref{tab:main}). Thus, linguistic triggers are comparatively weak carriers of action-level control in this setting.

The Text+Vision configuration closely mirrors Vision-only performance, maintaining 98.17\%-100\% ASR across budgets (Table~\ref{tab:main}). This reinforces that the visual channel dominates backdoor activation: once the visual trigger is present, adding a textual trigger provides no consistent improvement in ASR. At the lowest budget (0.31\%), ST for Text+Vision decreases modestly (95.50\% $\pm$ 6.08\%) relative to Vision-only (99.17\% $\pm$ 1.44\%), but at higher budgets ($\ge$ 1.25\%) stealthiness returns to 98\%--99\%. 

Overall, these results substantiate our central finding: in this VLA instantiation, action-level backdoor control is primarily mediated through the visual channel, remains highly effective even under extremely small poisoning budgets, and preserves both clean-task performance and precise temporal responsiveness. This confirms that safety-critical actions can be covertly and reliably manipulated with minimal collateral impact.

\begin{table}[t]
\centering
\caption{Ablation study on textual and visual trigger variants. We report ASR, ST, and RT (ms) averaged over three seeds (mean $\pm$ std).}
\resizebox{\columnwidth}{!}{%
\begin{tabular}{l l c c c}
\toprule
Modality & Variant & ASR \% & ST \% & RT (ms) \\
\midrule
\multirow{4}{*}{Text} 
& Non-natural token (\texttt{[sudo]}) & \(99.67 \pm 0.29\)  & \(98.50 \pm 1.32\) & \(7 \pm 1\) \\
& Natural connector (\texttt{now})    & \(99.67 \pm 0.29\)  & \(97.33 \pm 1.53\) & \(7 \pm 0\) \\
& Active adverb (\texttt{carefully})  & \(100.00 \pm 0.00\) & \(99.00 \pm 0.50\) & \(7 \pm 1\) \\
& Task-specific sentence              & \(99.83 \pm 0.29\)  & \(99.17 \pm 0.58\) & \(11 \pm 5\) \\
\midrule
\multirow{5}{*}{Vision}
& Circle, $1.0\times$, 1.0 & \(100.00 \pm 0.00\) & \(99.00 \pm 0.50\) & \(7 \pm 1\) \\
& Circle, $2.0\times$, 1.0 & \(98.33 \pm 2.89\)  & \(95.00 \pm 4.92\) & \(7 \pm 0\) \\
& Triangle, $1.0\times$, 1.0 & \(100.00 \pm 0.00\) & \(99.17 \pm 0.58\) & \(7 \pm 1\) \\
& Circle, $1.0\times$, 0.5 & \(99.50 \pm 0.50\)  & \(97.83 \pm 0.58\) & \(7 \pm 0\) \\
& Circle, $1.0\times$, 0.2 & \(100.00 \pm 0.00\) & \(98.83 \pm 0.76\) & \(8 \pm 1\) \\
\bottomrule
\end{tabular}}
\label{tab:abl-trigger}
\end{table}

\noindent\textbf{Trigger variants.}
Across diverse textual forms and moderate visual appearance variations (shape, scale, and opacity), the attack remains largely invariant: ASR consistently stays near 100\%, and ST remains high, with only a modest reduction in ST observed for the $2\times$ visual-trigger scale (Table~\ref{tab:abl-trigger}). RT remains consistently low across variants, indicating reliable and temporally precise overriding of the targeted action.

\subsection{Robustness to Trigger Mismatch}

We further evaluate robustness under deployment-time trigger mismatches using two representative backdoored models (Vision-only and Text+Vision). At inference time, we vary two factors independently: (i) the textual trigger (presence and surface form), and (ii) the visual trigger (appearance parameters and spatial placement). This setup isolates the stability of the implanted backdoor under moderate train--test discrepancies, reflecting realistic deployment variations.

\begin{table}[t]
\centering
\caption{Robustness under inference-time trigger mismatches. We report ASR and RT for Vision-only and Text+Vision backdoored models. }
\resizebox{\columnwidth}{!}{%
\begin{tabular}{l cc cc}
\toprule
\multirow{2}{*}{Mismatch case}
& \multicolumn{2}{c}{Text+Vision backdoored}
& \multicolumn{2}{c}{Vision-only backdoored} \\
\cmidrule(lr){2-3}\cmidrule(lr){4-5}
& ASR \% & RT (ms) & ASR \% & RT (ms) \\
\midrule
Default (circle, $1\times$, 1.0, default; text=\texttt{carefully})
& 100.0 & 8  & 99.5 & 7 \\
Opacity change (0.2; others default)
& 100.0 & 8  & 99.5 & 7 \\
Shape change (triangle; others default)
& 100.0 & 8  & 99.5 & 7 \\
Scale change ($2\times$; others default)
& 100.0 & 8  & 99.5 & 7 \\
Text variant (text=\texttt{now}; others default)
& 100.0 & 8  & 99.5 & 7 \\
Remove text trigger (text=none; others default)
& 100.0 & 8  & 99.5 & 7 \\
Remove visual trigger (visual trigger removed)
& \textbf{1.0} & 21 & \textbf{0.0} & -- \\
Position shift: center
& \textbf{31.0} & 23 & \textbf{0.5} & 4 \\
Position shift: bottom-right
& 100.0 & 11 & \textbf{85.5} & 19 \\
\bottomrule
\end{tabular}}
\label{tab:robust-merge}
\end{table}

\begin{table}[t]
\centering
\caption{Cross-suite zero-shot transfer on LIBERO-Goal using models fine-tuned on LIBERO-Spatial (OFT). $N_{\text{valid}}$ denotes the number of evaluable episodes, and ASR is computed conditionally on these episodes. }
\label{tab:libero_goal_transfer_asr_rt}
\begin{tabular}{lccc}
\toprule
Modality & $\mathbf{N_{\text{valid}}}$ & ASR (\%) & RT (ms) \\
\midrule
Text-only   & 138 & 0.72  & 14 \\
Vision-only & 134 & 96.27 & 7  \\
Text+Vision & 110 & 99.09 & 9  \\
\bottomrule
\end{tabular}
\end{table}

\noindent\textbf{Remarks.}
Each configuration is evaluated over 200 episodes. An episode contributes to $N_{\text{valid}}$ only if the rollout reaches the triggerable state, ensuring that the trigger can be applied and the target action meaningfully assessed. Episodes that never satisfy this precondition are excluded from ASR computation, preventing task execution failures (i.e., never entering the triggerable state) from being conflated with backdoor failures (i.e., trigger present but failing to induce the target action).

The results in Table~\ref{tab:robust-merge} characterize deployment-time trigger mismatches for the two backdoored models. For both Text+Vision and Vision-only poisoning, modifying the textual trigger form (e.g., \texttt{carefully} to \texttt{now}) has negligible effect on ASR; even removing the text trigger preserves near-100\% ASR. This indicates that, in our setting, the backdoor is largely insensitive to inference-time language cues. In contrast, removing the visual trigger collapses the attack (ASR $\approx$ 0--1\%, Table~\ref{tab:robust-merge}), confirming that the backdoor signal is primarily mediated through the visual channel.

Within the visual modality, moderate appearance-level perturbations (shape, scale, opacity) leave ASR essentially unchanged, demonstrating robustness to visual variations. However, spatial relocation exposes a sharp generalization boundary: moving the trigger to an unseen position (e.g., the image center) substantially degrades ASR, and even a milder shift to the bottom-right corner noticeably weakens Vision-only performance. These findings reinforce that while the visual signal dominates activation, its effectiveness depends critically on spatial consistency with poisoning-time placement.

\subsection{Cross-suite Zero-Shot Transfer}

We further evaluate cross-suite generalization in a zero-shot setting: models are fine-tuned (OFT) on \textit{LIBERO-Spatial} and directly evaluated on \textit{LIBERO-Goal} without any target-suite adaptation. 

As shown in Table~\ref{tab:libero_goal_transfer_asr_rt}, transfer performance exhibits strong modality dependence. Vision-only and Text+Vision backdoors retain high ASR on the unseen suite (96.27\% and 99.09\%, respectively), whereas the Text-only backdoor largely fails (0.72\%). This disparity indicates that transferable triggering is primarily mediated by stable and reusable visual cues, while purely textual triggers are more sensitive to distributional shifts, such as changes in instruction templates or language grounding across suites.

Importantly, once activated, the backdoor maintains low reaction latency (RT = 7--9\,ms for Vision-only and Text+Vision, approximately 3--5 control steps, computed over successful episodes). This temporal precision further supports the action-level backdoor hypothesis: the trigger induces a rapid and targeted override of a specific action, rather than causing broad policy degradation. Together, these results reinforce that visually anchored action-level backdoors can generalize across task suites while preserving precise, time-critical control.


\subsection{Real-world Experiment}

We further validate \textbf{DropVLA} in a real-world setting using a 7-DoF Franka Emika arm, with $\pi_0$-fast~\cite{pi0} as the backbone policy. Due to practical constraints of real-robot experimentation, we evaluate a single physical trigger instantiation: a blue cube placed at the designated location used during backdoor fine-tuning. Unlike simulation, the trigger object remains physically static in the scene, while its image-plane coordinates vary as the robot moves and the camera viewpoint changes. This setup directly probes the location-robustness dimension identified in our simulation studies, particularly sensitivity to camera-relative spatial shifts.

We evaluate three language-conditioned manipulation tasks, including ``put the blue cup on the plate" and two fried-chicken pick-and-place tasks (e.g., ``pick up the fried chicken into the rubbish can" and ``put the fried chicken on the plate"), under a poisoning rate of $\ p_{ep}= 4\%$. Across 200 real-world trials, the attack achieves a 20\% success rate under our physical-world metric definition. Although lower than simulation performance, this outcome is consistent with the spatial generalization degradation observed in our trigger relocation experiments. Importantly, it demonstrates that action-level backdoor effects persist beyond simulation and pose non-trivial practical risk in embodied deployments.

\section{Discussion}

\noindent\textbf{VLA-specific risk and the role of generalization.}
Action-level backdoors are particularly concerning in VLA policies because they operate within closed-loop control: even a brief deviation at a critical decision point can propagate into significant physical consequences. Unlike task-level hijacking, action-level attacks target reusable low-level actions (e.g., gripper commands or end-effector motions), making the trigger--action association inherently compositional across tasks. As a result, generalization is central to the threat. Without robustness to nuisance variation and task-context shifts, a trigger--action mapping cannot function as a reusable control mechanism in deployment settings.

\noindent\textbf{What our results imply.}
Our empirical findings indicate that the backdoor signal is predominantly mediated by the visual channel. Modifying or removing the language cue has negligible impact on ASR, whereas removing the visual trigger collapses the attack (Table~\ref{tab:robust-merge}). Within the visual modality, moderate appearance changes (shape, scale, opacity) transfer reliably, yet spatial relocation beyond the poisoning support induces a sharp drop in ASR, revealing a clear generalization boundary under single-location poisoning. Practically, this suggests that in-scene visual cues constitute a broad and plausible attack surface in embodied deployment. Accordingly, defenses should prioritize auditing and hardening visually conditioned execution of safety-critical actions.

\noindent\textbf{Limitations and future work.}
Our study focuses on a single target action and evaluates within the \texttt{LIBERO} benchmark family using OpenVLA-7B. Training triggers with location robustness is important in real-world deployment, since the trigger’s image-plane location can drift as the camera--robot relative pose changes over time. Extending this framework to multi-action trigger--action mappings with attacker-timed activation policies represents a natural next step. Moreover, systematically characterizing cross-task and cross-environment transfer under realistic poisoning budgets would further clarify the practical risk envelope of action-level backdoors.

\noindent\textbf{Practical implications and mitigations.}
Action-level backdoors are particularly hazardous because a trigger can override a \emph{safety-critical action} within a short control window, inducing irreversible state transitions (e.g., unintended object release) even when episode-level task success remains high. Effective defenses should therefore monitor the \emph{action interface} and the \emph{critical timesteps} at which such actions take effect. Concretely, potential mitigations include: (i) \emph{runtime gating} of safety-critical actions (e.g., \texttt{open\_gripper}) using lightweight short-horizon state or force-consistency checks, with delay-or-fallback rules when contextual inconsistency is detected; (ii) \emph{trigger-surface auditing} through time-local stress tests that inject structured in-scene visual perturbations (small shifts, crops, lighting changes) around critical windows and measure conditional action activation probabilities; and (iii) \emph{adaptation-time data hygiene}, including provenance tracking and similarity-based filtering to downweight or remove rare or near-duplicate episodes disproportionately containing safety-critical actions at critical decision points. These measures directly target the reusable trigger$\rightarrow$action mapping emphasized in our threat model, rather than relying solely on coarse episode-level outcomes.

\section{Conclusion}
\label{sec:conclusion}

In this work, we characterized a VLA-specific backdoor attack surface: \emph{action-level} targeted backdoors that hijack \emph{reusable} safety-critical action primitives at attacker-chosen decision points in closed-loop control. Under a realistic pipeline-black-box poisoning setting, we introduced \textbf{DropVLA} and showed that injecting an extremely small fraction of poisoned episodes can induce near-100\% targeted \texttt{open\_gripper} activation \emph{within a 0.05\,s post-onset reaction window} while preserving high trigger-free task success. 

Technically, DropVLA relies on a window-consistent supervision mechanism tailored to chunked VLA fine-tuning: after trigger onset we relabel a contiguous block of steps to maintain label consistency across overlapping training windows, enabling stable implantation of time-local action overrides under tiny poisoning budgets. Our modality and robustness analyses further revealed that backdoor activation is predominantly mediated by the visual channel: vision-only poisoning is consistently effective across budgets and transfers across suites, whereas text-only triggers are unstable under sparse poisoning and contribute little when combined with vision. Finally, we validated real-world feasibility on a 7-DoF Franka arm with $\pi_0$-fast; despite camera-relative motion that induces image-plane trigger drift, the attack attains a non-trivial 20.0\% success rate over 200 trials. Overall, these findings show that small-budget poisoning can implant precise, time-critical action-level overrides in VLA systems without observable degradation of nominal performance, underscoring the need for defenses that explicitly monitor, audit, and harden safety-critical action execution.

\noindent\textbf{Ethics Statement.}
This work studies backdoor vulnerabilities in Vision--Language--Action models to improve the safety of embodied AI systems. All experiments are conducted in controlled settings on public benchmarks and open-source models.  We follow responsible disclosure practices for any code release. We do not provide actionable instructions for deploying attacks on real robots.

\noindent\textbf{Code Availability.}
The code is publicly available at: \url{https://github.com/megaknight114/DropVLA}.

\clearpage

\bibliographystyle{IEEEtran}
\bibliography{refs}

\end{document}